          [2001/04/25 1.1 (PWD)]
\documentclass[a4paper,twocolumn]{esapub2005} 
\usepackage{times}
\usepackage{natbib}
\usepackage{graphicx}

\title{Balloon-borne gamma-ray polarimetry}
\author{Mark Pearce, for the PoGOLite Collaboration}
\affil{KTH, Department of Physics and The Oskar Klein Centre for Cosmoparticle Physics, AlbaNova University Centre, 10691 Stockholm, Sweden. Email: pearce@kth.se}

\begin{document}


\maketitle

\begin{abstract}
The physical processes postulated to explain the high-energy emission mechanisms of compact astrophysical sources often yield polarised soft gamma rays (X-rays). PoGOLite is a balloon-borne polarimeter operating in the 25-80 keV energy band. The polarisation of incident photons is reconstructed using Compton scattering and photoelectric absorption in an array of phoswich detector cells comprising plastic and BGO scintillators, surrounded by a BGO side anticoincidence shield. The polarimeter is aligned to observation targets using a custom attitude control system. The maiden balloon flight is scheduled for summer 2011 from the Esrange Space Centre with the Crab and Cygnus X-1 as the primary observational targets. 
\end{abstract}

\section{Introduction}
Polarised soft gamma-rays (X-rays) are expected from the high-energy processes at work within compact astrophysical objects such as pulsars, accreting black holes and jet-dominated active galaxies~\cite{overview}. The polarisation arises naturally for synchrotron radiation in large-scale ordered magnetic fields and for photons propagating through a strong magnetic field. Polarisation can also result from anisotropic Compton scattering. In all cases, the orientation of the polarisation plane is a powerful probe of the physical environment around compact astrophysical sources. Polarisation is also expected to provide valuable information regarding the processes underlying other astrophysical phenomena, such as solar flares~\cite{solar} and gamma-ray bursts~\cite{grb}.

Despite the wealth of sources accessible to polarisation measurements and the importance of these measurements, there has been only one successful mission with dedicated instrumentation. The Crab nebula was studied at 2.6 keV and 5.2 keV using a Bragg reflectometer flown on the OSO-8 satellite in 1976~\cite{weisskopf}. At 2.6 keV, the measured polarisation degree was (19.2$\pm$1.0)\%, with position angle (156.4$\pm$1.4)$^\circ$, while at 5.2 keV the measured values were (19.5$\pm$2.8)\% and (152.6$\pm$4.0)$^\circ$. The results are consistent with radiation arising from synchrotron processes. The polarimeter did not have sufficient effective area to allow significant measurements of the Crab pulsar.

Measurements making inventive use of instruments on-board the Integral satellite have reinvigorated the field of late, with polarisation measurements and limits reported for both the Crab and Cygnus X-1. The IBIS instrument was used to study the Crab in the 200-800 keV range using data collected between 2003 and 2007~\cite{crabIBIS}. No significant polarisation was reported for either pulsar peak. In the off-pulse region, $>$72\% polarisation and a position angle of (120.6$\pm$8.5)$^\circ$ was reported, changing to $>$88\% and (122.0$\pm$7.7)$^\circ$ if the bridge region was also included. As for OSO-8, the high polarisation degree indicates synchrotron emission from a well-ordered magnetic field region. The pulsar rotation axis has been estimated from X-ray observations to be (124.0$\pm$0.1)$^\circ$~\cite{MK39}, which is compatible with the position angle. Data from the SPI instrument collected between 2003 and 2006 and covering the energy range 100 keV-1 MeV has also been studied~\cite{crabSPI}. The polarisation degree was found to be (46$\pm$10)\% with a position angle of (123.0$\pm$11)$^\circ$. The IBIS instrument has also been used to observe Cygnus X-1. For the energy range 250-400~keV, a polarisation degree of (67$\pm$30)\% was found with a position angle of (140$\pm$15)$^\circ$. This is at least 100$^\circ$ from the compact radio jet. Spectral modeling of the data suggests that the low energy data are consistent with emission dominated by Compton scattering on thermal electrons. The higher energy data which is, in contrast, strongly polarised may indicate that emission in the MeV region is related to the radio jet. An upper limit of 20\% polarisation was set for the energy range 400-2000~keV. It is important to note that the IBIS and SPI instruments were not designed for polarimetric measurements and that their response to polarised radiation was not studied prior to launch.  

\section{Measurement principle and polarimeter design}
\begin{figure}
\centering
\includegraphics[width=1 \linewidth]{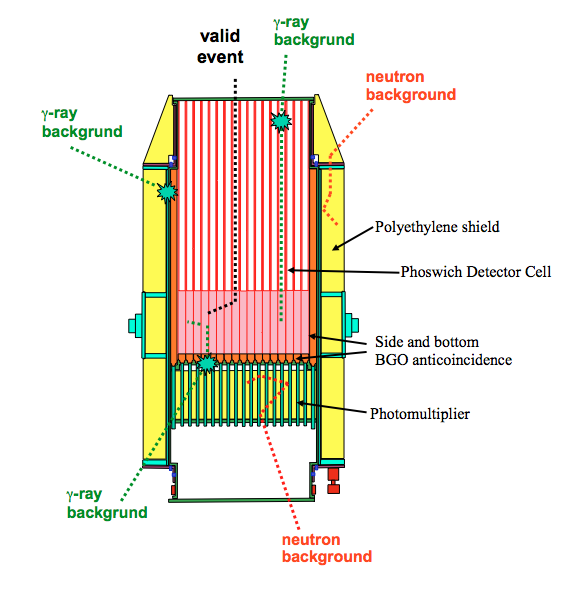}
\caption{A schematic overview of the PoGOLite polarimeter. The overall length is approximately 2~m. The main components are identified along with possible signal and background interactions. The volume below the photomultipliers houses data acquisition electronics, including a fluid-based cooling system. The polarimeter components are housed in a pressure vessel assembly which rotates around the viewing axis.
\label{fig:schematic}}
\end{figure}

PoGOLite~\cite{pogolite} is a balloon-borne soft gamma-ray (25-80 keV, arguably X-ray is a more correct term for photons in this energy range) polarimeter which is optimised for the study of point-like astrophysical objects such as pulsars and accreting black holes. 
A schematic of the PoGOLite instrument is shown in figure~\ref{fig:schematic}. Compton scattering and photoabsorption in an array of phoswich detector cells (PDC) with hexagonal cross-section, made of plastic and BGO scintillators and surrounded by a BGO side anticoincidence shield (SAS), is used to determine the polarisation of incident photons. The full-size PoGOLite instrument consists of 217 PDC units. Initially, a reduced volume 'pathfinder' instrument comprising 61 PDC units will be evaluated in flight. Each PDC is composed of a thin-walled tube (well) of slow plastic scintillator at the top (fluorescence decay time $\sim$280 ns), a solid rod of fast plastic scintillator (decay time $\sim$2 ns), and a short bismuth germanate oxide (BGO) crystal at the bottom (decay time $\sim$300 ns), all viewed by one photomultiplier tube (PMT). The wells serve as a charged particle anticoincidence, the fast scintillator rods as active photon detectors, and the bottom BGOs act as a lower anticoincidence. Each well is sheathed in thin layers of tin and lead foils to provide passive collimation. 

Gamma-rays entering within the field of view of the instrument (2 degrees $\times$ 2 degrees (FWHM), defined by the slow plastic scintillators) will hit one of the fast plastic scintillators and may be Compton scattered, with a probability that depends on the photon energy. The scattered photon may escape, be photoabsorbed in another detector cell, or undergo a second scattering. Electrons resulting from a photoabsorption will deposit their energy in the plastic scintillator and produce a signal at the PMT. A trigger based on the photoelectron energy deposit will initiate high speed waveform sampling of PMT outputs from all PDCs. Valid Compton scattering events will be selected from these waveforms after the completion of a flight. 

\begin{figure}[t]
\centering
\includegraphics[width=1 \linewidth]{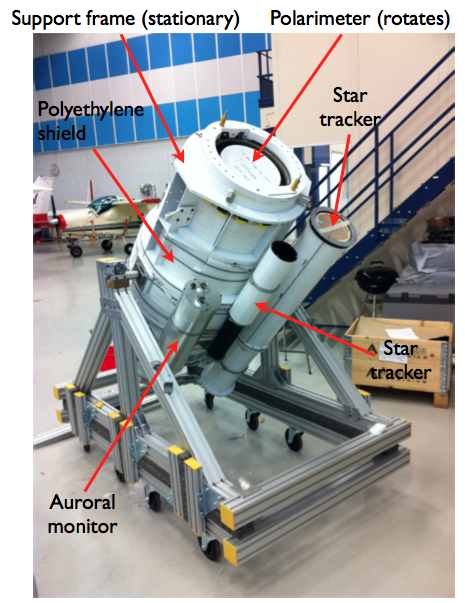}
\caption{The polarimeter assembly mounted in the transport wagon, prior to integration with the attitude control system. Two star trackers and an auroral monitor are attached to the body of the polarimeter. Due to the relatively high latitude (67$^\circ$N) of the pathfinder flight, there is a potential background from polarised X-rays arising from auroral bremmstrahlung at an altitude of 90-100~km~\cite{aurora}.
\label{fig:polarimeter}}
\end{figure}

The locations of the PDCs in which the Compton scatter and photoabsorption are detected determine the azimuthal Compton scattering angle. The geometry of the PDC arrangement limits the polar scattering angle to approximately (90$\pm$30) degrees, roughly orthogonal to the incident direction. Little of the energy of an incident gamma ray photon is lost at the Compton scattering site(s), while most of the energy is deposited at the photoabsorption site. This makes it straightforward to differentiate Compton scattering sites from photoabsorption sites. The azimuthal Compton scattering angles will be modulated by the polarisation of the photon. The polarisation plane can be derived from the azimuthal distribution of scattering angles. The degree of polarisation (\%) is determined from the ratio of the measured counting rate modulation around the azimuth to that predicted for a 100\% polarised beam (from simulations calibrated with experiments at polarised photon beams). The polarimeter assembly is placed inside a pressure vessel system which is in turn housed inside a structure which allows the polarimeter to rotate along the viewing axis. This allows instrument asymmetries to be systematically studied during flight. The pathfinder polarimeter is shown in figure~\ref{fig:polarimeter}.

A fluid-based cooling system is employed to manage the heat load ($\sim$200~W) generated inside the polarimeter. The heat originates from the DC/DC converters inside the tightly-packed array of 92 photomultipliers and data acquisition electronics. Fluid (Paratherm LR) is pumped around a cooling plate to which the PMTs are thermally bonded, and around the walls of the housing for the data acquisition electronics. Heat is dissipated through radiators which are directed at the cold sky during the flight. It is particularly important that the PMTs are operated close to room temperature in order to assure sensitivity to the small energy deposits expected from photons Compton scattering at the lower end of the PoGOLite energy range. Additional technical details regarding PoGOLite systems can be found elsewhere in these proceedings~\cite{miranda}.

\section{Performance studies}
\begin{figure}
\centering
\includegraphics[width=1 \linewidth]{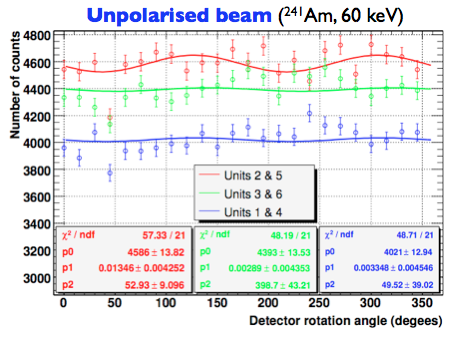}
\caption{Modulation curves obtained in the laboratory when illuminating the central detector cells of the pathfinder polarimeter shown in figure~\ref{fig:polarimeter} with unpolarised $\sim$60~keV photons from an $^{241}$Am radioactive source. During measurements the polarimeter is rotated around the viewing axis. Opposite pairs of PDCs are considered in order to reduce the effect of alignment systematics. No significant modulation is found.
\label{fig:unpolarised}}
\end{figure}

\begin{figure}
\centering
\includegraphics[width=1 \linewidth]{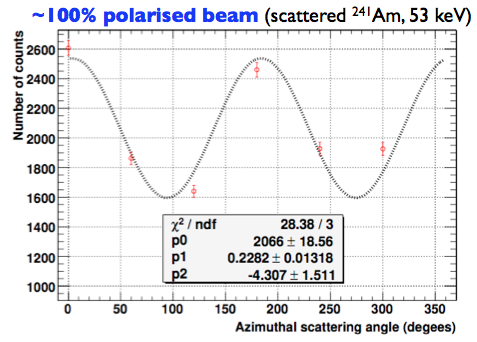}
\caption{The modulation curve obtained when illuminating the central detector cells of the pathfinder polarimeter with  $\sim$100\% polarised $\sim$53~keV photons from an $^{241}$Am radioactive source. A clear modulation signal is obtained.  
\label{fig:polarised}}
\end{figure}

The response of the pathfinder polarimeter has been studied in the laboratory using $\sim$60~keV photons from a $^{241}$Am radioactive source~\cite{mkthesis}. The response is quantified in terms of  the modulation factor which describes the amplitude of the counting rate asymmetries reconstructed in the polarimeter. The response to an unpolarised beam is shown in figure~\ref{fig:unpolarised}. The modulation curve is constructed for a pair-wise combination of detector cells in order to reduce the effect of alignment systematics. The resulting modulation curves are seen to be compatible with no modulation, which is  important given that polarisation is a positive definite measurement in the presence of unknown systematics. A polarised beam was generated by Compton scattering the $^{241}$Am beam through 90 degrees in a small plastic block. The resulting scattered beam will be approximately 100\% polarised with an energy of $\sim$53~keV. An example modulation curve obtained during this study is shown in figure~\ref{fig:polarised}. A modulation factor of (22.8$\pm$1.3)\% is reconstructed which is in line with expectations from Geant4 simulations and previous studies of a prototype polarimeter at a polarised synchrotron beam facility~\cite{mkthesis}, ~\cite{kek}. 


\begin{figure}[t]
\centering
\includegraphics[width=1 \linewidth]{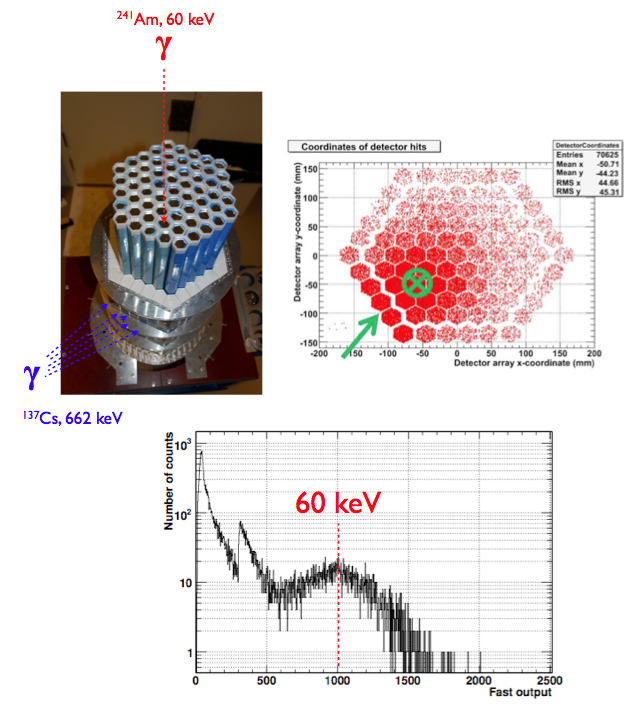}
\caption{Left: The irradiation geometry used to study gamma-ray background rejection. The source rates were chosen to be close to those expected in flight. Right: An cumulative event view during the irradiation study. The green cross shows the location of the $^{241}$Am 'signal' beam, while the green arrow shows the irradiation direction from the $^{137}$Cs background. The effect of the background is clearly seen in the PDCs. Bottom: The reconstructed photopeak from $^{241}$Am photons after background reduction through inspection of anticoincidence signals and kinematic requirements on scattering events.
\label{fig:137cs}}
\end{figure}

Figure~\ref{fig:polarised} shows the background contributions for a Crab observation by the PoGOLite pathfinder. These results are from Geant4 simulations of the pathfinder instrument in a flight-like configuration. The minimum ionising particle signature in scintillators arising from charged cosmic-ray backgrounds is readily reduced using a simple pulse height analysis. A background due to neutrons (mostly albedo) generated by cosmic-ray interactions with the earth's atmosphere is seen to dominate. Neutrons in the energy range 500~keV to 2~MeV are the main component of the false trigger rate, with a flux of 0.7 neutrons s$^{-1}$cm$^{-2}$ predicted below 10 MeV.
For this reason a polyethylene shield (with thickness 15~cm around the scattering scintillators) has been added to the PoGOLite design. This reduces the neutron background by an order of magnitude. In order to measure residual background during flight, a dedicated neutron detector comprising a 5~mm thick LiCaAlF$_6$ crystal with 2\% Eu doping complemented with BGO anticoincidence is mounted in the vicinity of the fast scattering scintillators~\cite{licaf}. The $^{6}$Li in the crystal has a large capture cross-section (940 barn) for thermal neutrons, giving sensitivity to neutrons with energies $<$10~MeV. 
The photon background arises from both atmospheric and galactic sources. This background is suppressed due to the narrow field-of-view provided by the PDC design and the segmented anticoincidence system. A demonstration of this background rejection is provided in figure~\ref{fig:137cs}. The pathfinder instrument was illuminated from the side (at the level of the scattering scintillators) with 662~keV photons from a $^{137}$Cs radioactive source. Simultaneously $\sim$60~keV 'signal' photons from a $^{241}$Am radioactive source were directed into a PDC. Source rates were chosen to mimic those expected from the Crab and backgrounds. As seen in the figure, the 60 keV photopeak could be reconstructed from the PDC data by applying pulse shape discrimination and simple kinematic considerations.

\begin{figure}
\centering
\includegraphics[width=1 \linewidth]{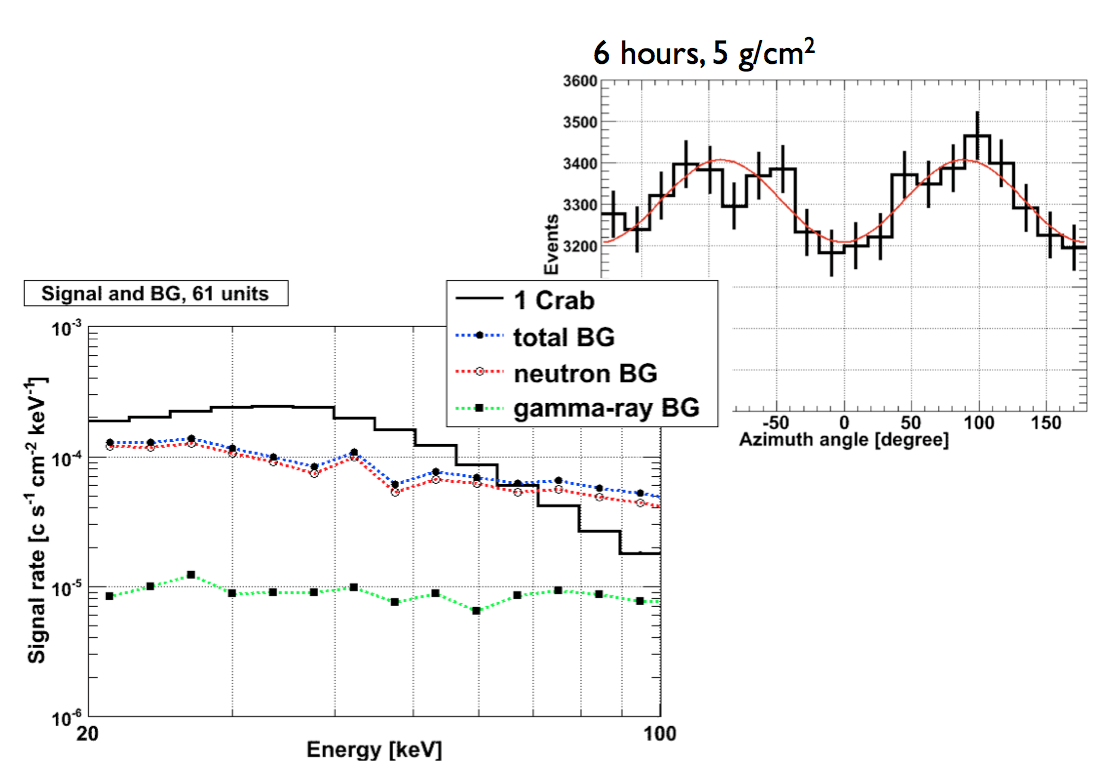}
\caption{Left: Signal and background energy spectra obtained from Geant4 simulations of the pathfinder polarimeter. The atmospheric neutron background is found to dominate. Right: The modulation curve expected from a 6~hour long observation of the Crab with 5~g/cm$^2$ of residual atmosphere.
\label{fig:background}}
\end{figure}

\section{Attitude control system (ACS)}
\begin{figure}
\centering
\includegraphics[width=1 \linewidth]{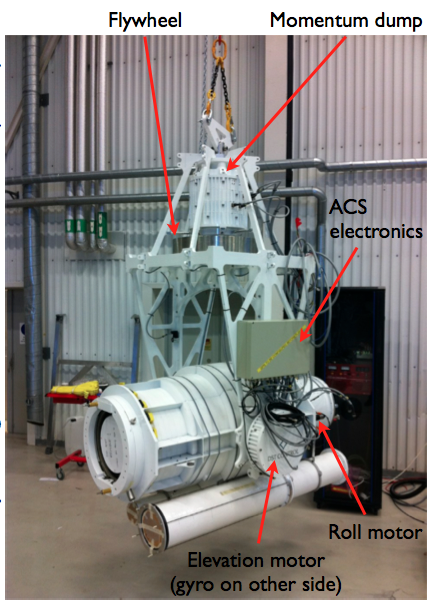}
\caption{The PoGOLite attitude control system. The polarimeter is mounted in a gimbal assembly where custom direct drive torque motors act on the polarimeter elevation axle. The balloon flight train connects to the gimbal assembly through a combined momentum dump/flywheel system which is used for azimuthal positioning. 
\label{fig:acs}}
\end{figure}

The narrow field-of-view of PoGOLite means that it is optimized for point-like sources. The ACS keeps the viewing axis of the polarimeter aligned to targets of interest as they track across the sky and also compensates for local perturbations such as flight-train torsion and changing stratospheric winds. Pointing to within $\sim$5\% of the polarimeter field-of-view (corresponding to $\sim$0.1$^\circ$) is required to secure a minimum detectable polarisation of better than 10\% for a 1 Crab source during the pathfinder flight for a 6~hour long observation. The ACS is discussed elsewhere in these proceedings~\cite{jes}. The polarimeter is shown mounted in the ACS gimbal assembly in figure~\ref{fig:acs}. Custom torque motors act directly on the polarimeter elevation axis. Azimuthal positioning is achieved with a flywheel assembly which connects to the flight train through a momentum dump motor which allows angular momentum stored in the flywheel to be reset upon saturation. Control signals to the motor systems are generated by a real-time computer system which monitors the attitude sensors, comprising a differential GPS system, a micromechanical accelerator/gyroscope package, angular encoders, an inclinometer, and a magnetometer. These primary attitude sensors are augmented by two star trackers. One star tracker is developed from a design successfully used on the HERO, HEFT and BLAST missions~\cite{stm}. This tracker has a field-of-view of 2.57$^\circ \times $1.92$^\circ$ and exploits a combination of (a) a relatively long focal length (300 mm) to minimise the solid angle subtended by CCD pixels; (b) extensive Aeroglaze-coated (infra-red absorbant) baffling; and (c) a red filter to suppress scattered radiation. Ground-based tests~\cite{cmb} have shown that stars down to 10$^{\mathrm{th}}$ magnitude can be resolved assuming the background light conditions at 40~km reported in~\cite{cmb91}. The other star tracker is a more compact design with in-house optics providing a field-of-view of 5.0$^\circ \times $3.7$^\circ$. Both trackers require a stabilized gondola to operate (provided by the differential GPS system and gyroscopes) and can either operate in a star pattern matching mode (providing relatively slow, but absolute position fixes to the ACS) or can be commanded to lock onto a bright star providing ACS updates at a rate of order 1 Hz. The latter case is foreseen to be the primary operating mode during flight.

\section{Gondola and ancillary systems}
\begin{figure}
\centering
\includegraphics[width=1 \linewidth]{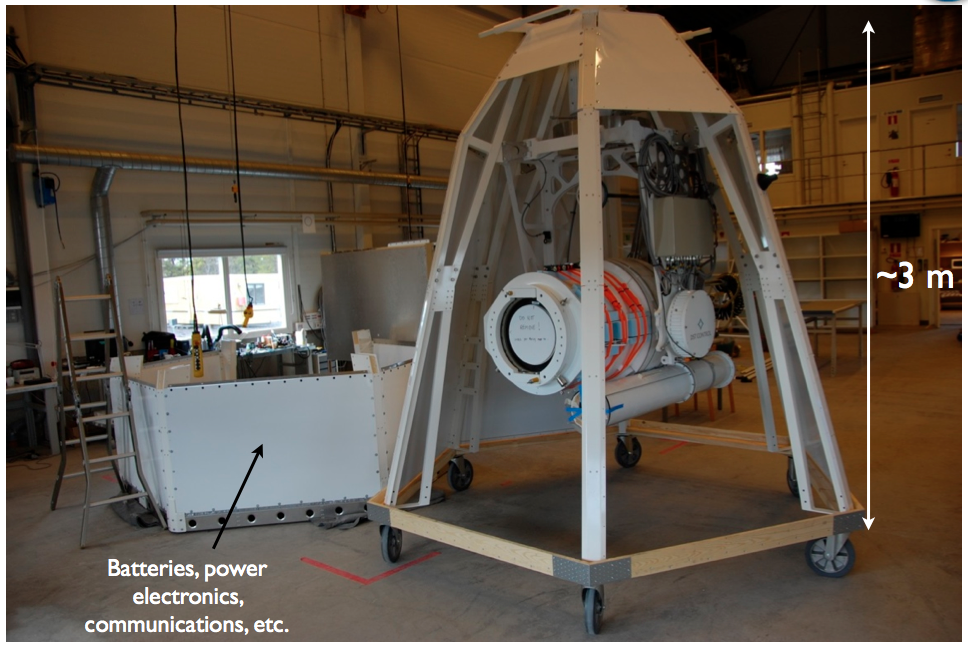}
\caption{The polarimeter mounted inside the upper part of the gondola during integration.  
\label{fig:gondola1}}
\end{figure}

\begin{figure}
\centering
\includegraphics[width=1 \linewidth]{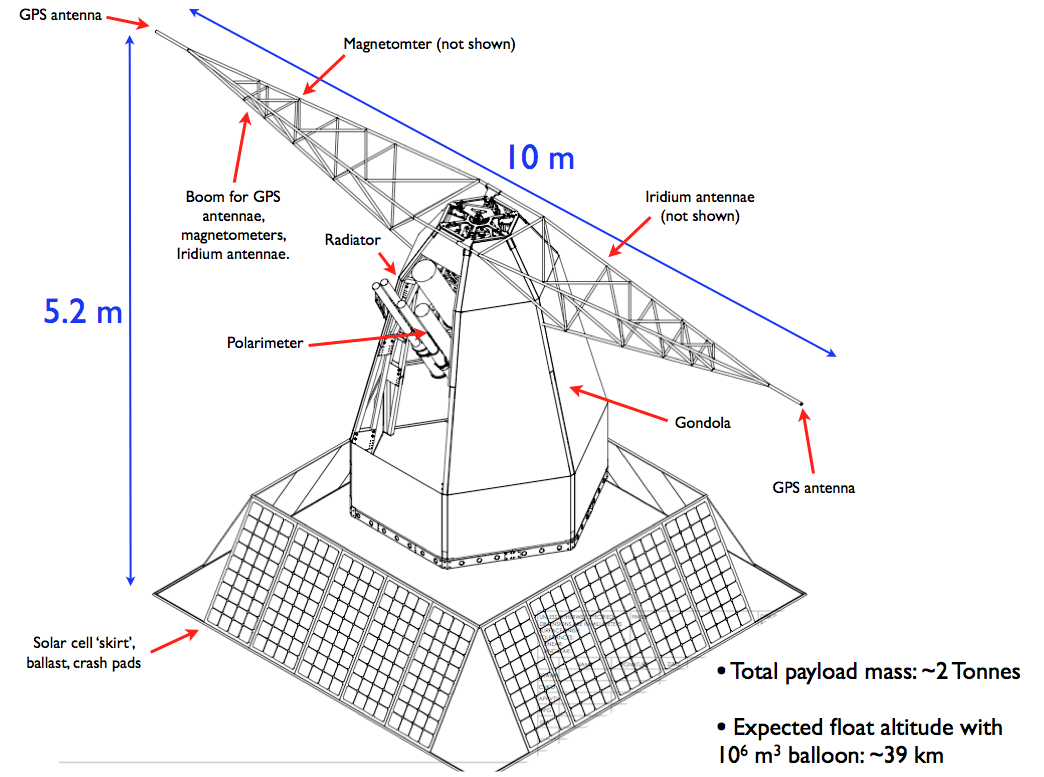}
\caption{A schematic overview of the complete gondola assembly. The gondola is mounted upon a solar cell skirt which also houses crash pads and ballast hoppers. The gondola is crowned by a lightweight boom construction which is 10~m long in order to provide the desired dGPS pointing accuracy.
\label{fig:gondola2}}
\end{figure}

As shown in figures~\ref{fig:gondola1} and~\ref{fig:gondola2}, the polarimeter and ACS are installed in a two part gondola, designed by SSC Esrange. The upper part of the gondola is connected to the ACS frame and also provides a mounting point for the cooling system radiators and pump. The lower section houses batteries, power control electronics and communications equipment. The gondola frame is covered in lightweight honeycomb panels which enhance the structural rigidity and help protect the polarimeter from damage during landing. Two glass-fibre booms are attached to the upper gondola. The booms span 10 m and have a GPS antenna mounted at each end. The antenna separation provides the required baseline for the differential GPS system alone to meet the $\sim$0.1$^\circ$ pointing accuracy requirement. Other GPS antennae for Esrange flight systems, Iridium antennae for over-the-horizon communications and magnetometers for the ACS and auroral monitor are also mounted on the booms. Beneath the lower gondola, a four-sided 'skirt' of solar panels (each side comprising 5 panels, with overall dimensions (3.5 $\times$ 1.5)~m is mounted along with landing crash pads, ballast hoppers and E-Link communication antennae. The gondola stands approximately 5.2~m tall and has a flight-ready mass of approximately 2~Tonnes, including ballast.


\section{Flight plan}
The POGOLite pathfinder was originally scheduled to fly in mid-August 2010 during stratospheric wind turn-around conditions, and a 24 hour long flight was foreseen. This flight was cancelled in the aftermath of the failed launch of the NCT payload at Alice Springs in April 2010. Thereafter, the PoGOLite flight was rescheduled for July 2011 with a circumpolar trajectory proposed, pending fly-over agreement between the Swedish and Russian governments (the balloon will be cut down over Western Canada after $\sim$5 days if fly-over permission are not granted by Russia). A circumpolar flight presents excellent scientific prospects due to the long ($\sim$ 20 days) flight time allowing repeated target observations. The primary targets for the PoGOLite Pathfinder are the Crab (nebula and pulsar) and Cygnus X-1 (if in the hard state). Targets of opportunity during the flight may also be considered. A July flight places stricter requirements on the ACS compared to an August flight since the Sun is closer to the Crab. The separation angle is $\sim$15 degrees in early July and  $\sim$65 degrees in mid-August. It may therefore not be possible to use the star trackers during the initial phase of a July flight and there will be more reliance on the dGPS system. Observations of Cygnus X-1 are possible during either period. As well as making observations of science targets, significant time will be spent making background observations of 'blank sky' regions in order to confirm the response to unpolarised radiation.


\end{document}